\documentclass[sigconf]{acmart}

\usepackage{enumitem}
\usepackage{algorithm}
\usepackage{algpseudocode}
\usepackage{booktabs} 
\usepackage{multirow}

\copyrightyear{2019}
\acmYear{2019} 

\acmConference[AdKDD]{}{}{}
\acmBooktitle{}
\acmPrice{}
\acmDOI{}
\acmISBN{}

\usepackage{balance}

\copyrightyear{2019} 
\acmYear{2019} 
\setcopyright{none}
\acmConference[AdKDD '19]{AdKDD 2019}{August 5, 2019}{Anchorage, AK}
\acmBooktitle{AdKDD 2019, August 5, 2019, Anchorage, AK}
\begin{document}
\title{Learning from Multi-User Activity Trails for B2B Ad Targeting}

\author{
{Shaunak Mishra, Jelena Gligorijevic, and Narayan Bhamidipati}
}
\affiliation{Yahoo Research}
\email{shaunakm,jelenas,narayanb@verizonmedia.com}




\begin{abstract}
Online purchase decisions in organizations can go through a complex journey
with multiple agents involved in the decision making process.
Depending on the product being purchased,
and the organizational structure, the process may involve
employees who first conduct market research, and then influence decision makers who place the online
purchase order. In such cases, the online activity trail of a single individual in the organization may only provide
partial information for predicting purchases (conversions). To refine conversion prediction for
business-to-business (B2B) products using online activity trails, we introduce the notion of \textit{relevant} users in 
an organization with respect to a given B2B advertiser, and leverage the collective activity trails of such relevant users
to predict conversions. In particular, our notion of relevant users is tied to a seed list of relevant activities for a B2B advertiser, and we propose a method using distributed activity representations to build such a seed list.
Experiments using data from Yahoo Gemini demonstrate that the proposed methods
can improve conversion prediction AUC by $8.8\%$, and provide an interpretable advertiser specific list of activities useful for B2B ad targeting.
\end{abstract}

\maketitle

\section{Introduction} \label{sec:introduction}
\begin{figure}[!htb]
\centering
  \includegraphics[width=0.95 \columnwidth]{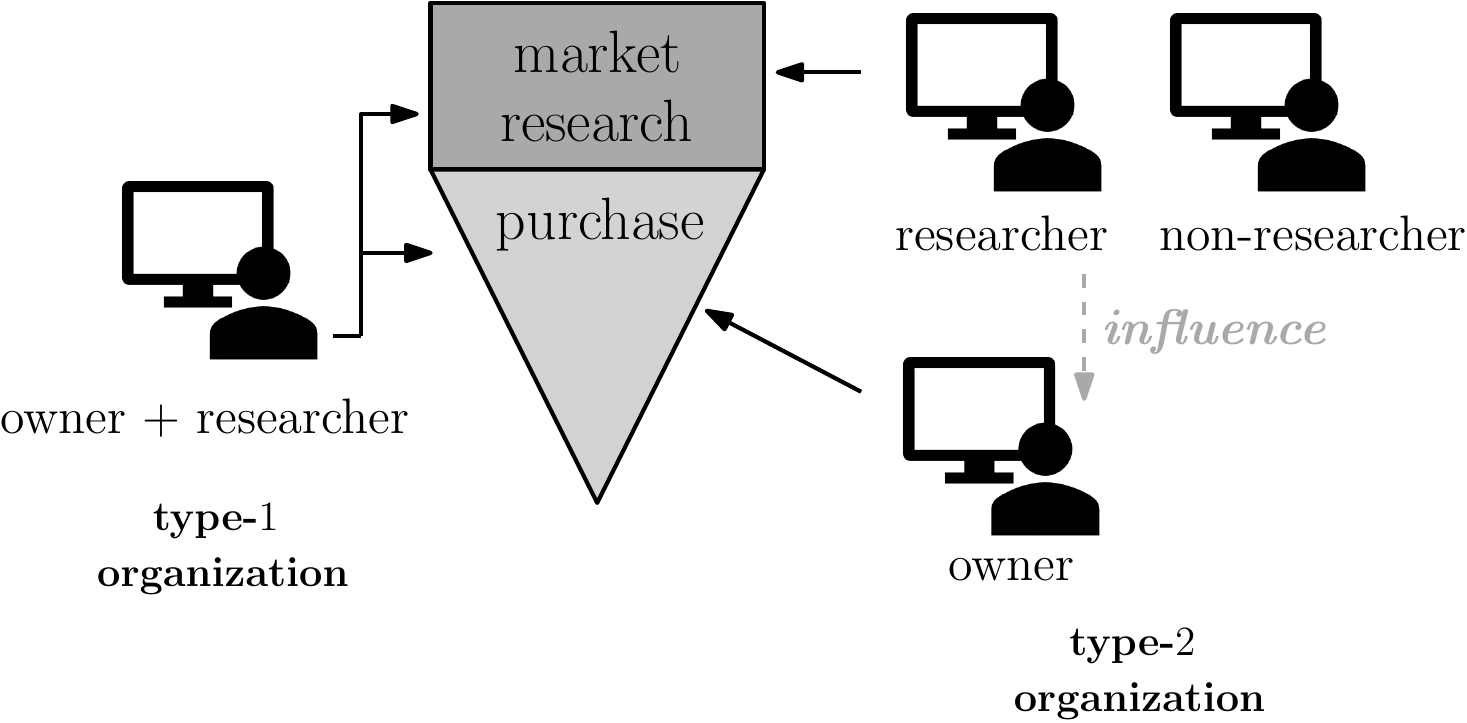}
  \caption{Example of type-$1$ and type-$2$ organizations vis-a-vis the B2B purchase (conversion) funnel. In a type-$1$ organization, a single user does the market research and converts, while in type-$2$, the market research and conversion activities are split across users in the same organization.}
  \label{fig:pull_figure}
\end{figure}

Even before the dawn of online advertising, understanding purchase decisions in organizations was considered a complex topic \cite{webster}; with the B2B interaction opportunities added by online advertising platforms, the complexity has only increased over time \cite{mckinsey_b2b}. A fundamental factor contributing to such complexity is the presence of multiple agents involved in various \textit{stages} of the decision process \cite{mckinsey_b2b,wiki_b2b}. For example, in the purchase funnel terminology \cite{mckinsey,jansen2011bidding}, some employees (researchers) in an organization 
may perform market research (upper funnel activities), and then pass on relevant information to the decision makers (owners) who may just place the purchase order (lower funnel activities).
There might also be smaller organizations where a single employee does all the market research and places the purchase order. Figure~\ref{fig:pull_figure} illustrates such an example, and introduces the notion of type-$1$ and type-$2$ organizations.
Intuitively, if an advertising platform \textit{knows} a priori which online users are owners + researchers (in type-$1$),
or just researchers (in type-$2$), it can efficiently target such individuals with relevant B2B ads \cite{linkedin_kdd2016}.
Such knowledge can be derived from multiple sources including: (i) declared user profiles in professional social networks (\emph{e.g.} LinkedIn), and (ii) data from customer relationship management (CRM) tools, \emph{e.g.}, Salesforce. The impact of using such proprietary user profile knowledge can be gauged by the following statistic: the
projected spending on B2B digital advertising in the US in 2019 is $6$ billion USD,
and ad platforms based on professional social networks account for over one-fifth of it \cite{b2b_2019_spending}. \\

In contrast to the scenarios mentioned above, in this paper, we consider a setup where such user profiles are not known a priori to an advertising platform. We assume that the advertising platform has access to online user activity trails (\emph{e.g.}, search queries, site visits, ad interactions), and has the ability to cluster users in the same geographic neighborhood at the granularity of a household or commercial establishment\footnote{Such user clustering can be based on deterministic or probabilistic methods for cross-device identity management \cite{neufeld_crossdevice, learning_to_rank_crossdevice}.}.
This setup is fairly common in (business-to-consumer) B2C ad platforms, where click and conversion prediction models \cite{mappi_cikm,google_FTRL} are trained on user activity trails to target users with relevant ads.
In this paper, we propose methods to refine such B2C conversion prediction models for B2B ads by leveraging the activities performed by users in a cluster as opposed to using only a single user's activity trail. Using the notion of type-$1$ and type-$2$ organizations, the following toy example explains the fundamental problem encountered by a B2C conversion model when naively used for predicting B2B conversions.

\paragraph*{Toy example:}
Consider a situation, where for a B2B advertiser, user activity trails from only type-$2$ organizations are present in the training data for a conversion prediction model. 
Each user is marked with a unique user-id, \emph{i.e.}, an owner, researcher, and non-researcher in an organization have separate user-ids. Furthermore, the activity distribution
is such that the owner does not do any market research, and if a purchase decision is made, the purchase (conversion) activity appears only in the owner's activity trail. In comparison, market research activities relevant to the conversion event, are only present in the activity trails of researchers. Given such training data, if a conversion prediction model is trained solely on basis of user activity trails (\emph{i.e.}, labeling only the converted owner's as positive and the rest as negative), it will have poor performance since it completely misses out on activities predictive of conversion. On the other hand, if the training data had only type-$1$ organizations, the conversion model can easily learn predictive activities since the market research activities
and the conversion activity occur in the same converter trail. Learning such predictive activities
is crucial for targeting users with relevant B2B ads.\\

As the above toy example suggests, the presence of type-$2$ organizations can degrade the performance of a conversion model working on just user activity trails.
In practice, depending on the B2B advertiser, the training data can be a mix of trails from type-$1$ and type-$2$ organizations. It is plausible that for an advertiser relevant for smaller businesses, the fraction of conversions from type-$1$ organizations may be dominant, whereas for an advertiser relevant for both small and large businesses, the conversions may span both type-$1$
and type-$2$ organizations in significant proportions; we assume that the distribution of such organizations is not known a priori for a given B2B advertiser.
An intuitive approach to get around the problems introduced by type-$2$ organizations
can be to use the trails of all users in an organization (user cluster) who have done activities \textit{relevant}
to the B2B advertiser. We build on top of this intuition, and introduce scalable approaches for leveraging such \textit{multi-user} trails for the task of B2B conversion prediction, as well as inferring
activities relevant for targeting ads of a given B2B advertiser.
Our main contributions can be summarized as follows.
\begin{enumerate}
\item For a given B2B advertiser, we introduce a notion of \textit{relevant} users within a user cluster.
We show that augmenting trails of all users in a cluster by adding activities of such relevant users can significantly improve user conversion prediction ($\sim8.8\%$ AUC lift in our experiments using data from Yahoo Gemini). The relevant users are based on a seed list of relevant activities for the given B2B advertiser.

\item For generating the seed list of relevant activities for a given B2B advertiser, we propose starting with an initial list and expanding it using a set expansion method based on distributed activity representations (referred to as activity2vec in the same spirit as word2vec \cite{w2v}). We validate the efficacy of the expansion method
using a logistic regression based B2B conversion prediction model. Such relevant activities can be directly used to create segments for B2B ad targeting.

\item We provide an information theoretic justification behind augmenting trails of users in a cluster with activity trails of relevant users within the same cluster. In particular, we show how such augmentation reduces the conditional entropy (uncertainty) in the conversion event random variable.
\end{enumerate}
The remainder of the paper is organized as follows.
Section~\ref{sec:related} covers related work,
and Section~\ref{sec:data} describes the data sources used in this paper.
Section~\ref{sec:method} goes over the proposed method, and Section~\ref{sec:info_theory} deals with the information theoretic justification behind augmenting user trails.
Section~\ref{sec:results} covers experimental results,
and we decribe conclusions in Section~\ref{sec:discussion}.

\section{Related work} \label{sec:related}
In this section, we cover related work on online advertising (Section~\ref{sec:ads}),
purchase behavior modeling in B2B and B2C setups (Section~\ref{sec:purchase}),
activity embeddings (Section~\ref{sec:embeddings}), and
online identity management approaches (Section~\ref{sec:id_management}).

\subsection{Online advertising} \label{sec:ads}
Brands (advertisers) typically signup with ad platforms (\emph{e.g.}, Google Ads, Facebook Ads, Yahoo Gemini)
to show their ads to online users.
As a part of setting up online ad campaigns in an ad platform, advertisers
may create one or more creatives (ad text and images) to target relevant audience (\emph{i.e.}, ad groups),
and for each ad group they specify a bid \cite{broder2008computational}.
During the auction for ad serving \cite{broder2008computational,mappi_cikm},
the bid may be used in conjunction with the predicted click through rate (CTR) and the predicted conversion rate (CVR).
Such CTR and CVR prediction models are trained based on historical user data \cite{google_FTRL,mappi_cikm,gligorijevicdeeply}
with click and conversion labels, and are crucial for advertisers to target relevant users.
In large scale advertising setups, logistic regression (LR) models have been successfully used \cite{google_FTRL,mappi_cikm} for CTR and CVR prediction. Recently, deep learning models have also been introduced in this context, \emph{e.g.}, deep residual networks \cite{deep_crossing} and sequential models like recurrent neural networks (RNNs)
\cite{jelena_sigir2018,
cui2018modelling}.
In this paper, we use an LR model for conversion prediction (details in Section~\ref{sec:method}), but our methods can be extended to
sequential models, \emph{e.g.}, RNNs.

\subsection{B2C and B2B purchase behavior models} \label{sec:purchase}
In general, an online user may go through various stages of the purchase funnel \cite{funnelwiki,mckinsey}
(\emph{e.g.}, unaware, aware, interest, consideration, intent) before purchasing from an advertiser.
In the B2C context, where a single individual's journey can be mapped through the purchase funnel,
the funnel structure can be leveraged
for ad targeting \cite{jansen2011bidding,geminix_kdd}.
In a similar spirit, B2B marketing literature includes purchase funnel studies which indicate
that multiple users in an organization may go through such funnel stages before the organization decides to purchase
\cite{mckinsey_b2b,webster}. In the context of professional social
networks there has been work on identifying key decision makers within an organization (via declared user profiles) for the purpose of
B2B ad targeting \cite{linkedin_kdd2016}. In the B2B marketing industry, identifying such \textit{leads} \cite{duncan_kdd2015}
and key decision makers in an organization is widely seen as an effective approach for B2B marketing \cite{linkedin_kdd2016}.
Compared to prior work in professional social networks (where users explicitly declare their roles in an organization),
we do not focus on identifying generic decision makers in an organization. In particular, we focus on leveraging online activity trails (\emph{e.g.}, search queries, site visits) to identify users relevant to the purchase decision for a given B2B advertiser.

\subsection{Activity embeddings and corpus-based set expansion}  \label{sec:embeddings}
Online users perform a variety of activities including reading articles, search queries,
purchasing items, visiting websites, and interacting with ads.
Across the activities done by a user, there can be richer semantic relationships,
and many such activities can be editorially mapped to stages in the purchase funnel \cite{jansen2011bidding}.
However, discovering such semantic relationship across activities with web scale data is non-trivial.
For example, a user planning a trip to a theme park in Orlando, could look for hotels in Orlando, check about weather,
and query for the theme park deals prior to the trip; hence, such events occurring in the user's trail are not isolated events,
but are related to the activity of buying the theme park's ticket.
Understanding such activity relationships can help in identifying users likely to click and convert on the theme park ads in the above example.
In the past, similarities between activities of one particular kind: either search \cite{search2vec} or purchase data from email extractions \cite{prod2vec} have been derived using embedding models constructed similar to word2vec \cite{w2v}.
In this paper, we take a similar approach to identify multi-modal activities \textit{relevant} to a given B2B conversion via activity2vec embeddings (details in Section~\ref{sec:act2vec}).
In addition, we iteratively refining such a seed list of relevant activities taking inspiration from
corpus based set expansion methods \cite{set_expan_first,set_expan} used to expand a small list of entities, while maintaining the same semantic class.
In particular, we propose an iterative (activity) seed list expansion method which tries to preserve the
semantic relationship of activities with respect to a B2B conversion.

\subsection{Online identity management} \label{sec:id_management}
With the proliferation of mobile devices, users switching between desktop and mobile devices
(\emph{e.g.}, laptop, phone, tablet) have made it challenging for ad platforms to track a user's online history.
Approaches for cross-device identity management are still evolving, and it is a topic of great importance in the online advertising industry. Currently there exist both deterministic as well as probabilistic algorithms for such cross device identity management \cite{neufeld_crossdevice, learning_to_rank_crossdevice}, and major ad platforms typically use proprietary methods.
In addition, there are deterministic ways to identify commercial IP addresses.
\section{Data} \label{sec:data}
In this paper, we leverage two data sources as described below.
\paragraph*{User activity trails:}
We use user activity trails data provided by Verizon Media. This includes online activities done in chronological order by a user. The activities are derived from heterogeneous sources, \emph{e.g.}, Yahoo Search, Yahoo Gemini ad interactions,
and viewing content on other publishers associated with Yahoo. The representation of an activity comprises
of an activity ID, time stamp, the type (\emph{e.g.}, search, content view), and a raw description of the activity (\emph{e.g.}, the exact search query for search activities) after stripping personally identifiable information.
In total, there are more than $3$ billion unique activities in our data spanning over $100$ million anonymized users.

\paragraph*{User ID $\rightarrow$ Cluster ID map:}
We use user ID to cluster ID maps determined by proprietary identity management algorithms at Verizon Media.
Such clusters represent groups of users deemed to belong to the same household or organization. In our data set, we had over $92$ million unique cluster IDs.

\section{Methodology} \label{sec:method}
In this section, we first give a high level overview of our proposed approach in Section~\ref{sec:overview}.
This is followed by details on the conversion prediction model (Section~\ref{sec:relevant_conversion_model}), and seed list generation (Section~\ref{sec:seed_list}).

\subsection{Overview} \label{sec:overview}
We introduce the notion of a \textit{relevant} activities seed list for a B2B advertiser. This is
a list of online activities which are expected to be performed before a B2B conversion; both owners and researchers spanning type-$1$ and type-$2$ organizations can perform such activities prior to a conversion event (seed list generation details in Section~\ref{sec:seed_list}).
Given a seed list, we identify \textit{relevant} users in each (user) cluster\footnote{We assume that the ad platform has access to the cluster information based on cross device identity management and IP addresses.} as follows: in each cluster, we find users who have performed at least one relevant activity in the seed list, and refer to them as \textit{relevant} users. Having identified relevant users, we train a conversion prediction model to estimate:
\begin{align}\label{eq:cnv_pred}
user \;conv. \;probability\; = \; \mathbb{P}\left(conv  \bigg |   trail_{u}, \{trail_r\}_{r \in \mathcal{C}_u} \right ),
\end{align}
where $trail_u$ denotes a user's trail of activities\footnote{The user activity trail is the sequence of user activities up to the time the conversion prediction is being made.
},
$\mathcal{C}_u$ denotes the cluster(s) the user is present in,
$\{trail_r\}$ is the collection of trails of all relevant users in $\mathcal{C}_u$. For a given B2B advertiser,
the conversion model is fed information from others users in the cluster who have done activities relevant to the B2B conversion event. In other words, for
a user's conversion prediction,
the activity trail of the user is augmented with the activities of other relevant users in the same cluster.
For scalability in terms of augmented trail lengths, and to control the noise injected while augmenting trails, we augment trails only via relevant users in a cluster.
The details of the conversion prediction model are covered in Section~\ref{sec:relevant_conversion_model}.
In summary, the conversion prediction model is designed towards better user conversion prediction,
and the seed list can be used to identify relevant users involved in the conversion decision process
(business owners and researchers) in an organization. Identifying such users can be helpful for B2B ad targeting segments.

\subsection{Conversion prediction model} \label{sec:relevant_conversion_model}
As mentioned above in Section~\ref{sec:overview}, we train an advertiser specific conversion prediction model using a user's activity trail, augmented with activity trails of all relevant users in the cluster(s) associated with the user.
We use an LR model for the purpose of conversion prediction; a one-hot encoded feature vector of activities from the augmented trails is used as input.
Although the choice of LR was motivated by scalability \cite{mappi_cikm,google_FTRL}, our approach can be extended to more sophisticated models like RNNs.
For training and testing the prediction model, label generation is done such that users who converted are given a positive label, and the users who did not convert are given a negative label (regardless of their cluster which may or may not contain a converter).

\subsection{Seed list generation algorithm} \label{sec:seed_list}
Intuitively, the relevant activities seed list should include activities which can identify researchers from both type-$1$
and type-$2$ organizations for a given B2B advertiser. For example, the seed list can cover activities like
visiting the advertiser's website, search queries for the advertiser's product or a competing brand's offerings in the same product category. In a web scale setup with billions of unique activities (as in our setup), obtaining such an \textit{interpretable} seed list can be challenging if the conversions are sparse, and if type-$2$ organizations dominate the data. For example, a naive conversion rate (per activity) based method may not be reliable in such a setup. To get around such challenges,
we propose the following two step process: (i) create a small yet interpretable initial seed list, and (ii) iteratively expand this initial list using activity2vec embeddings (described below) to add \text{similar} activities.
We provide below details on initial seed list generation (Section~\ref{sec:initial_seed_list}),
activity2vec (Section~\ref{sec:act2vec}), and seed list expansion (Section~\ref{sec:expansion}).

\subsubsection{Initial seed list} \label{sec:initial_seed_list}
For an initial seed list $S_{initial}$, we select the top $k$ activities by conversion rate for the given B2B advertiser.
The conversion rate for an activity $a_i$ is
defined as the ratio of count of users who did $a_i$, and converted within a time window (\emph{e.g.}, $2$ months) over the count of users who did $a_i$. The choice of $k$ can be adjusted to do editorial curation within time constraints, \emph{e.g.}, a few hundred activities can be reviewed by an editor in a matter of hours. The editor(s) can also add a few obvious activities like visiting web sites of the advertiser
to the intial seed list. However, this conversion rate based method suffers
from noise arising from sparse conversions, and the presence of type-$2$ organizations.
Editorial curation is done to remove such noise from the initial seed list, and passed on to the activity2vec based expansion described below.

\subsubsection{Activity2vec} \label{sec:act2vec}
Similar to the word2vec embedding model \cite{w2v}, we train a
skip-gram based embedding model of activities (activity2vec), and obtain a $300$ dimensional embedding for each activity in our data. For training, each user's chronological trail of online activities is treated as a document, and activity sessions are treated as sentences within the document.
Each sentence comprises of the activities (in chronological order)
done within the session, where a session is defined as a sequence of consecutive
activities that have inter-activity time gaps of less than $30$ minutes.
Using locality-sensitive hashing (LSH), for each activity, the top nearest neighbors in the activity2vec space
can be obtained in a scalable manner. 
 
\subsubsection{Expansion algorithm} \label{sec:expansion}
Given an initial seed list $S_{initial}$ of relevant activities (as discussed in Section~\ref{sec:initial_seed_list}),
we add related activities using the activity2vec based expansion as decribed in Algorithm~\ref{alg:seedlist_expansion_act2vec}.
In particular, we assume a train-test data set for conversion prediction as described in Section~\ref{sec:relevant_conversion_model}, and $conv\_prediction\_AUC(S)$ returns the test AUC (area under ROC curve) metric, when the conversion model ingests augmented user trails, \emph{i.e.}, a user trail is augmented with the collective trails of all relevant users (determined by seed list $S$) in the cluster.
In addition, $neighbors \left (S , \mathcal{V}, \Delta_{sim}, \Delta_{nbr} \right )$, returns
the list of activities $\in \mathcal{V}$ and $\not \in S$ which have activity2vec cosine similarity
greater than a threshold $\Delta_{sim}$ for at least $\Delta_{nbr}$ activities in $S$. In simple words,
the algorithm iteratively expands the current seed list with new activities which have high similarity with a lot of current seed list activities. The expansion algorithm terminates when conversion prediction using relevant users' trails does not improve in terms of AUC after adding new activities (which may be just noise).
\begin{algorithm}[H] 
\caption{\bf B2B seed list expansion using activity2vec}\label{alg:seedlist_expansion_act2vec}
\begin{algorithmic}[1] 
\item initialize $\mathcal{V} = set\;of\;all\;activities$
\item initialize $S_0=S_{initial}$
\item initialize $i = 1$
\item initialize $AUC_0 =  conv\_prediction\_AUC(S_0)$
\item initialize $stopping\_criteria = FALSE$
\While {$stopping\_criteria = FALSE$}{
\State $N_i = neighbors \left (S_{i-1} , \mathcal{V}, \Delta_{sim}, \Delta_{nbr} \right )$
\State $AUC_i  = conv\_prediction\_AUC(S_{i-1} \cup N_i)$
 \If {$ AUC_{i} >  AUC_{i-1} + \epsilon $ }
\State  $ S_i = S_{i-1 } \cup N_i $
\Else 
\State $stopping\_criteria = TRUE$
  \EndIf
\State $i = i +1$
\EndWhile}
\item final seed list = $S_{i-1}$
\end{algorithmic}
\end{algorithm}  

\section{Information theoretic analysis} \label{sec:info_theory}
In this section, we provide an information theoretic justification for augmenting the trails of users in a cluster with the trails of other relevant users in the cluster.
In particular, we analyze a simple setup where there exists a relevant activity which can \textit{perfectly} predict conversion of the owner in an organization.
In other words, an owner  converts if and only if the owner or a researcher in the same organization performs the relevant activity.
Even in this simple setup, it is plausible (\emph{e.g.}, in a type-$2$ organization)
that the relevant activity does not occur in the owner's activity trail, and still the owner converts under the influence of a researcher (who is not labeled as a researcher a priori but is in the same organization) performing the relevant activity.
Hence, just looking at an individual's activity trail for the presence of the relevant activity may not be sufficient to perfectly predict conversion. Intuitively, augmenting the owner's activity trail with the researcher's activity trail can help conversion prediction in this situation, and we theoretically prove how such augmentation can reduce the entropy (information theoretic uncertainty) \cite{cover_and_thomas} associated with the conversion event.
We use the terms cluster and organization interchangeably in the analysis described below. We first describe additional notation in Section~\ref{sec:info_notation},
and then analyze type-$2$ organizations in Section~\ref{sec:info_type2}.
\subsection{Analysis setup} \label{sec:info_notation}
Consider a homogeneous setup with the following notation:
\begin{itemize}
\item $k$: number of organizations,
\item $r$: researchers per organization,
\item $n$: non-researchers per organization,
\item $d$: owners per organization ($=1$),
\item $s $: total users in an organization ($= n + r + 1$),
\item $p_{u}$: conversion probability of a user, and
\item $p_{o}$: conversion probability of an organization ($=sp_{u}$),
\item $C$: binary random variable indicating user conversion ($C=1$ if user converts, else $C=0$), 
\item $R$: binary random variable indicating relevant activity in user trail ($R=1$ if relevant activity present in trail, else $R=0$), 
\end{itemize}
where all organizations are of the same size ($=s$), and an organization is said to convert if there exists a converter user belonging to the organization. As per the above notation,
on an average there are $kp_o$ organizations out of $k$ who convert.
Also, since each organization has $s$ users, out of which at most one (\emph{i.e.}, the owner) can have the \textit{user} conversion label, the relation $p_u = \frac{p_o}{s}$ holds.
In the remainder of this section, for simplicity, we will assume that, if and only if, a relevant activity is done by any researcher in an organization, the owner in the organization converts with probability $1$. Hence, if the organization was type-$1$, where the researcher is the owner, the following information theoretic relation would hold:
\begin{align}
I\left( C ; R \right) = H \left( C \right)  - H \left ( C | R   \right)  \stackrel{(a)}= H \left( C \right), \nonumber 
\end{align}
where $I\left( C ; R \right)$ denotes the mutual information \cite{cover_and_thomas} between the user conversion indicator $C$, and the relevant activity indicator $R$.
The entropy\footnote{For a binary random variable $X$, the entropy \cite{cover_and_thomas} is defined as $H(X) = \sum_{x \in {0,1} }  - p(X=x)log\left( p(X=x)\right)$, where $p(\cdot)$ is the probability mass function.}
of $C$ is denoted by $H(C)$, the conditional entropy of $C$ given $R$ is denoted by $H(C|R)$,
and step (a) follows from the assumption that for a type-$1$ organization $C=1$ iff $R=1$, \emph{i.e.}, given $R$ there is no uncertainty left in the value of $C$.
The mutual information is representative of how well the observation variable $R$
can predict the outcome variable $C$; as shown above, for a type-$1$ organization with the above assumptions, $R$ can predict $C$ perfectly. \\

Following the definition of type-$2$ organizations introduced in Section~\ref{sec:introduction}, for our analysis we assume that all $k$ organizations are of type-$2$, \emph{i.e.}, if they convert, the relevant activities are done by a researcher, and the act of conversion is performed by the owner.
From the mutual information calculation above, one can observe that increasing the mutual information between $C$ and $R$ boils down to reducing
the conditional entropy of $C$ given $R$; this motivates the analysis for conditional entropy as described below.

\subsection{Conditional entropy analysis for type-$2$ organizations}  \label{sec:info_type2}
\begin{figure}[!htb]
\centering
  \includegraphics[width=0.65 \columnwidth]{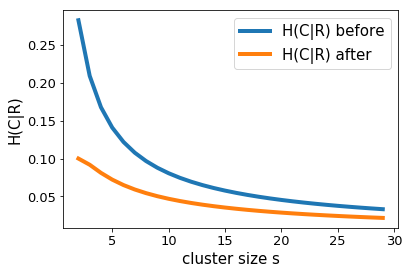}
  \caption{Comparison of $H(C|R)$ before and after trail augmentation for $p_0=0.1$ and $r=1$ across a range of $s$.} 
  \label{fig:info_analysis}
\end{figure}

We first analyze a toy example, and then generalize it as follows.
\begin{example}
Consider a data set with two type-$2$ organizations (clusters) as shown in Figure~\ref{fig:info_example}.
The conditional entropy $H(C|R)$ can be computed before and after trail augmentation as follows.
\begin{figure}[!htb]
\centering
  \includegraphics[width=0.95 \columnwidth]{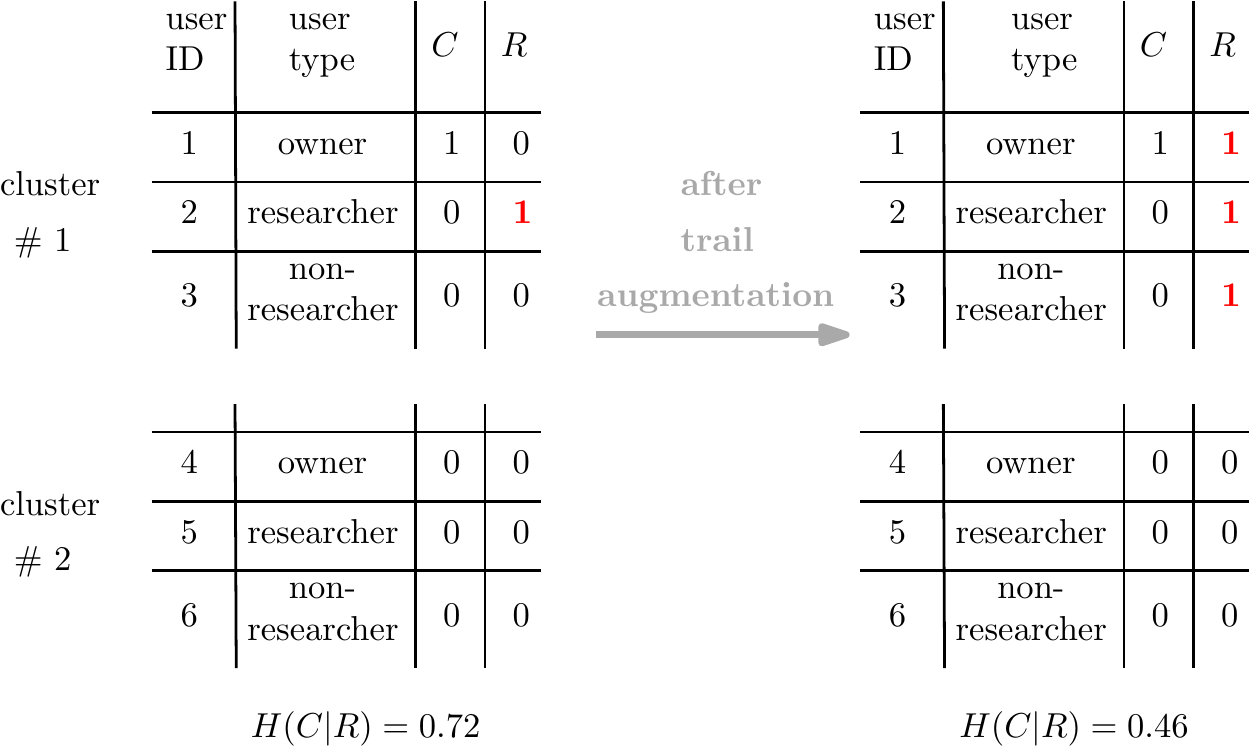}
  \caption{Toy example with two clusters such that $n=1, r=1, k=2, p_{o} = \frac{1}{2}, p_{u} = \frac{1}{6}$. Only the researcher in cluster $1$ has a relevant activity ($R=1$), and the owner in the cluster converts. After trail augmentation, all users in cluster $1$ have $R=1$. The conditional entropy $H(C|R)$ decreases after trail augmentation (equals $0.46$ versus $0.72$ before augmentation).} 
  \label{fig:info_example}
\end{figure}
\paragraph*{$H(C|R)$ before augmentation:}
\begin{align}
H\left (C|R \right) & = H\left(C|R=0\right) \mathbb{P}\left(R=0 \right) + H\left(C|R=1\right) \mathbb{P}\left(R=1 \right) \nonumber \\
& =  H\left ( \mathcal{B} \left (\frac{1}{5} \right ) \right )  \times \frac{5}{6} + 0 \times \frac{1}{6} = 0.72,  \nonumber
\end{align}
where $ \mathcal{B} \left (\frac{1}{5} \right )$ denotes a Bernoulli distribution with mean $\frac{1}{5}$. Also, $H\left(C|R=1\right)$$= 0$ holds since $C=0$ when $R=1$
in the example.
\paragraph*{$H(C|R)$ after augmentation:}
\begin{align}
H\left (C|R \right) & = H\left(C|R=0\right) \mathbb{P}\left(R=0 \right) + H\left(C|R=1\right) \mathbb{P}\left(R=1 \right) \nonumber \\
& =  0 \times \frac{1}{2} +   H\left ( \mathcal{B} \left (\frac{1}{3} \right ) \right )  \times \frac{1}{2} = 0.46 \nonumber 
\end{align}
\end{example}
CIearly the conditional entropy is lower after augmentation in the above example; this implies better conversion prediction given $R$. In fact, this advantage holds for a broader regime
of parameters (\emph{i.e.}, $r, p_o, s$) as derived below. For a general setup, the $H(C|R)$ before augmentation is:
\begin{align}
H(C|R) &= H\left(C|R=0\right) \mathbb{P}\left(R=0 \right) + H\left(C|R=1\right) \mathbb{P}\left(R=1 \right) \nonumber \\
& = H\left(C|R=0\right) \mathbb{P}\left(R=0 \right)
= H\left(  \mathcal{B} \left (   \frac{  p_0  }{s - r p_0}   \right )   \right)   \times \left (1- \frac{r p_o}{s} \right ), \nonumber
\end{align} 
and after augmentation it becomes:
\begin{align}
H(C|R) &= H\left(C|R=0\right) \mathbb{P}\left(R=0 \right) + H\left(C|R=1\right) \mathbb{P}\left(R=1 \right) \nonumber \\
& = 0 + H\left(C|R=1\right) \mathbb{P}\left(R=1 \right)
 = H\left( \mathcal{B} \left (\frac{1}{s} \right )  \right)  \times p_o.  \nonumber 
\end{align}
Based on the above derivation,
Figure~\ref{fig:info_analysis} shows the $H(C|R)$ comparison for a choice of $p_o=0.1$, and $r=1$ across a range of $s$. Clearly, $H(C|R)$ is lower after augmentation, but the gap decreases as $s$
increases.
This is intuitive since for fixed $p_o$, as $s\rightarrow \infty$, $p_u \rightarrow 0$, and hence $H(C) \rightarrow 0$. 
By definition \cite{cover_and_thomas}, $H(C) \leq H(C|R) $,  so as $s\rightarrow \infty$, $H(C|R) \rightarrow 0$. Hence, as Figure~\ref{fig:info_analysis} suggests, the benefit of trail augmentation is significant for \textit{smaller} organizations in particular.

\section{Experimental Results} \label{sec:results}


To evaluate the proposed seed list expansion approach, as well as the idea of augmenting user trails (via relevant users) for the task of conversion prediction,
we carried out (offline) conversion prediction experiments based on a B2B advertiser's data  from Yahoo Gemini.
In particular, we studied how different iterations of seed list expansion help the proposed B2B conversion prediction model, when compared to a baseline B2C model
where no trail augmentation is done (\emph{i.e.}, seed list $S=\phi$).
In other words, for each user, the B2C model uses only the user's activity trail,
while the B2B model uses the user's augmented activity trail, where the augmentation is based on a seed list (as described in Section~\ref{sec:relevant_conversion_model}).\\

\begin{figure}[h!]
	\centering
	\includegraphics[width=0.85 \columnwidth]{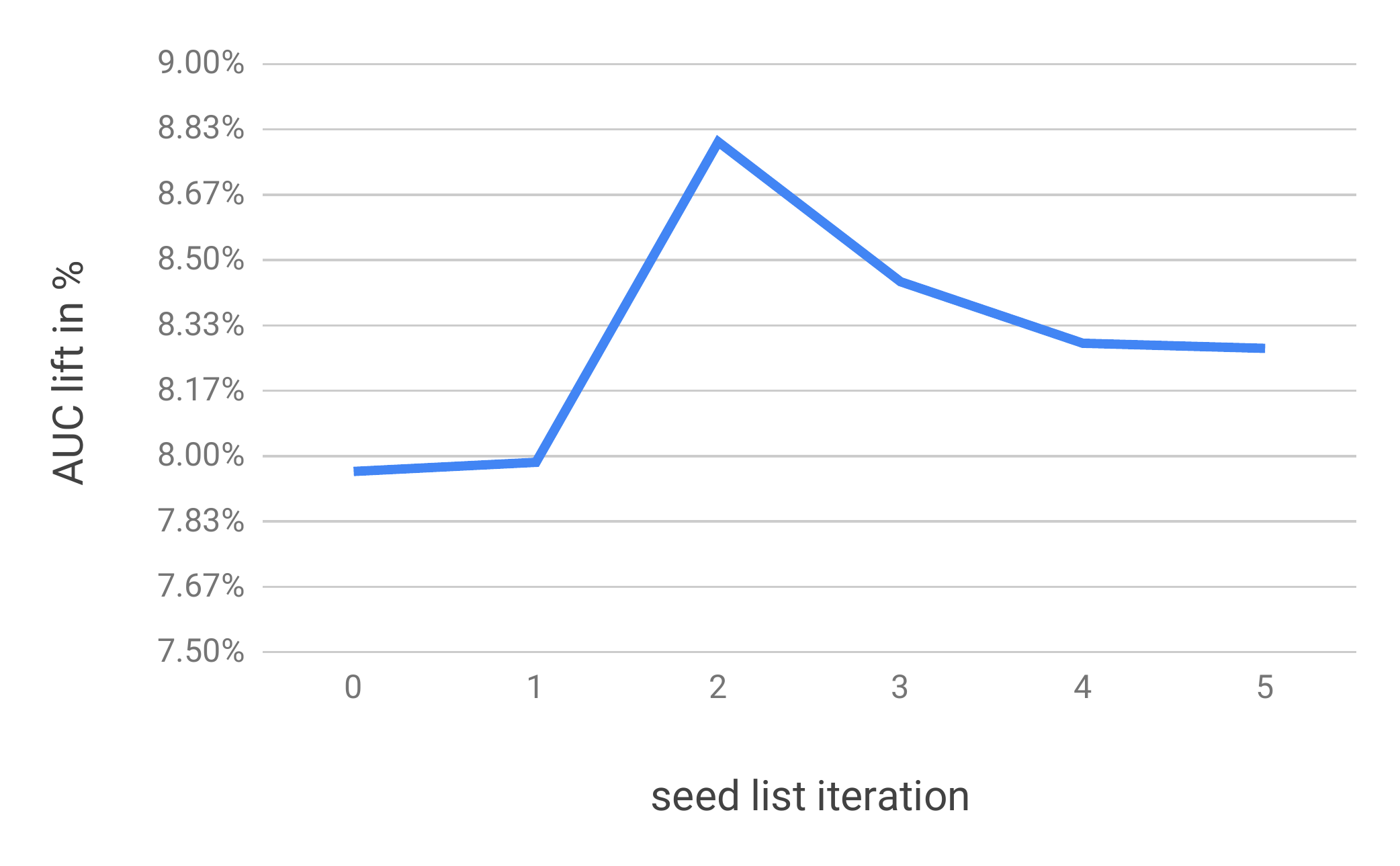}
	\caption{
		AUC lifts over B2C model (ingesting only user trails) for the proposed B2B model (using user trails plus relevant users from the same cluster) across seedlist iterations.}
	\label{fig:res_auc_lifts}
\end{figure}

Table~\ref{tab:auc_lifts} summarizes the conversion prediction (test) AUC lifts obtained for different versions of the seed list arising from iterations of the proposed
expansion algorithm. It also shows, for each iteration, the lifts in the number of activities in the expanded seed list  (compared to the initial seed list),
and the average count of relevant users (including converters) per converted cluster. As expected, these numbers are monotonically increasing.
The average number of relevant users (including converters) per cluster is observed to be close to one; this is partially related to the large percentage of single user clusters in the data set, and indicates that many organizations in our data set may be of type-$1$.
As shown, expanding the seed list with similar activities, marks more users in a cluster as relevant,
and the resultant trail augmentation demonstrates strong conversion prediction AUC lifts (peaking at iteration $2$).
\begin{table}[h!]
	\caption{ (1) AUC lifts from the proposed B2B conversion prediction model over B2C model  
	for $S_{initial}$ to $S_5$ iterations of seed list expansion. (2) Lifts in the number of activities in the seed lists compared to $S_{initial}$.
       (3) Average count of relevant users (including converters) in a converted clusters. 
} \label{tab:auc_lifts}
	\begin{tabular}{c||c|c|c}
		\textbf{Seed list} &                                     &                                              & \textbf{\# relevant users per} \\
		\textbf{iteration} & \multirow{-2}{*}{\textbf{AUC lift}} & \multirow{-2}{*}{\textbf{\#activities lift}} & \textbf{converter cluster}     \\ \hline
		
		\textbf{$S_{initial}$}        & 7.96\%                              & 
		-                                            & 1.241                          \\
		\textbf{$S_{1}$}        & 7.98\%                              & 5.24\%                                       & 1.278                          \\
		\textbf{$S_{2}$}        & \textbf{8.80\%}                     & 6.07\%                                       & 1.283                          \\
		\textbf{$S_{3}$}        & 8.44\%                              & 7.54\%                                       & 1.297                          \\
		\textbf{$S_{4}$}        & 8.29\%                              & 10.85\%                                      & 1.317                          \\
		\textbf{$S_{5}$}        & 8.27\%                              & 12.87\%                                      & 1.325                         

	\end{tabular}
\end{table}
A visualization of the AUC lifts is provided in Figure \ref{fig:res_auc_lifts}.
Guided by the AUC lifts, seed list expansion can be terminated after iteration $2
$, and the seed list $S_2$ can be used for relevant user detection and B2B ad targeting.

\section{Conclusion} \label{sec:discussion}
The proposed trail augmentation approach shows strong performance lifts for the B2B advertiser considered in our experiments. The information theoretic arguments in this paper support the approach for a wider regime,
making it is useful for a significant class of advertisers (for whom type-$2$ organizations are customers).
Also, the proposed seed list expansion provides a scalable method to infer relevant users for B2B ad targeting in an interpretable manner.

\bibliographystyle{abbrv}
\bibliography{literature}

\begin{thebibliography}{10}

\bibitem{wiki_b2b}
Business marketing.
\newblock \url{https://en.wikipedia.org/wiki/Business_marketing}.

\bibitem{b2b_2019_spending}
{US} {B2B} digital advertising trends.
\newblock
  \url{https://www.emarketer.com/content/us-b2b-digital-advertising-trends}.

\bibitem{funnelwiki}
Wikipedia: {P}urchase {F}unnel.
\newblock \url{https://en.wikipedia.org/wiki/Purchase_funnel}.

\bibitem{mappi_cikm}
N.~Bhamidipati, R.~Kant, S.~Mishra, and M.~Zhu.
\newblock A large scale prediction engine for app install clicks and
  conversions.
\newblock In {\em CIKM 2017}.

\bibitem{broder2008computational}
A.~Z. Broder.
\newblock Computational advertising.
\newblock In {\em SODA}, 2008.

\bibitem{mckinsey}
E.~Court, S.~Mulder, and O.~Vetvik.
\newblock The {C}onsumer {D}ecision {J}ourney.
\newblock {\em McKinsey Quarterly}, 2009.

\bibitem{cover_and_thomas}
T.~M. Cover and J.~A. Thomas.
\newblock {\em Elements of Information Theory}.
\newblock Wiley, 2006.

\bibitem{cui2018modelling}
Y.~Cui, R.~Tobossi, and O.~Vigouroux.
\newblock Modelling customer online behaviours with neural networks:
  applications to conversion prediction and advertising retargeting.
\newblock {\em arXiv preprint arXiv:1804.07669}, 2018.

\bibitem{duncan_kdd2015}
B.~A. Duncan and C.~P. Elkan.
\newblock Probabilistic modeling of a sales funnel to prioritize leads.
\newblock KDD '15.

\bibitem{jelena_sigir2018}
D.~Gligorijevic, J.~Gligorijevic, A.~Raghuveer, M.~Grbovic, and Z.~Obradovic.
\newblock Modeling mobile user actions for purchase recommendation using deep
  memory networks.
\newblock SIGIR '18.

\bibitem{gligorijevicdeeply}
J.~Gligorijevic, D.~Gligorijevic, I.~Stojkovic, X.~Bai, A.~Goyal, and
  Z.~Obradovic.
\newblock Deeply supervised model for click-through rate prediction in
  sponsored search.
\newblock {\em Data Mining and Knowledge Discovery}, 2019.

\bibitem{search2vec}
M.~Grbovic, N.~Djuric, V.~Radosavljevic, F.~Silvestri, R.~Baeza-Yates, A.~Feng,
  E.~Ordentlich, L.~Yang, and G.~Owens.
\newblock Scalable semantic matching of queries to ads in sponsored search
  advertising.
\newblock SIGIR '16.

\bibitem{prod2vec}
M.~Grbovic, V.~Radosavljevic, N.~Djuric, N.~Bhamidipati, J.~Savla, V.~Bhagwan,
  and D.~Sharp.
\newblock E-commerce in your inbox: Product recommendations at scale.
\newblock {\em KDD}, 2015.

\bibitem{jansen2011bidding}
B.~J. Jansen and S.~Schuster.
\newblock Bidding on the buying funnel for sponsored search and keyword
  advertising.
\newblock {\em Journal of Electronic Commerce Research}, 2011.

\bibitem{mckinsey_b2b}
O.~Lingqvist, C.~Plotkin, and J.~Stanley.
\newblock Do you really understand how your business customers buy?
\newblock Mc{K}insey Quarterly February 2015.

\bibitem{google_FTRL}
H.~B. McMahan, G.~Holt, D.~Sculley, M.~Young, D.~Ebner, J.~Grady, L.~Nie,
  T.~Phillips, E.~Davydov, D.~Golovin, S.~Chikkerur, D.~Liu, M.~Wattenberg,
  A.~M. Hrafnkelsson, T.~Boulos, and J.~Kubica.
\newblock Ad click prediction: a view from the trenches.
\newblock KDD 2013.

\bibitem{w2v}
T.~Mikolov, I.~Sutskever, K.~Chen, G.~Corrado, and J.~Dean.
\newblock Distributed representations of words and phrases and their
  compositionality.
\newblock NIPS'13, 2013.

\bibitem{neufeld_crossdevice}
E.~Neufeld.
\newblock Cross-device and cross-channel identity measurement issues and
  guidelines.
\newblock {\em Journal of Advertising Research}, 57(1):109--117, 2017.

\bibitem{deep_crossing}
Y.~Shan, T.~R. Hoens, J.~Jiao, H.~Wang, D.~Yu, and J.~C. Mao.
\newblock Deep crossing: Web-scale modeling without manually crafted
  combinatorial features.
\newblock KDD 2016.

\bibitem{set_expan}
J.~Shen, Z.~Wu, D.~Lei, J.~Shang, X.~Ren, and J.~Han.
\newblock Setexpan: Corpus-based set expansion via context feature selection
  and rank ensemble.
\newblock In {\em ECML/PKDD}, 2017.

\bibitem{learning_to_rank_crossdevice}
J.~{Walthers}.
\newblock Learning to rank for cross-device identification.
\newblock In {\em 2015 IEEE International Conference on Data Mining Workshop
  (ICDMW)}, Nov 2015.

\bibitem{set_expan_first}
R.~C. {Wang} and W.~W. {Cohen}.
\newblock Language-independent set expansion of named entities using the web.
\newblock In {\em ICDM)}, 2007.

\bibitem{webster}
F.~E. Webster and Y.~Wind.
\newblock A general model for understanding organizational buying behavior.
\newblock {\em Journal of Marketing}, 36(2):12--19, 1972.

\bibitem{linkedin_kdd2016}
S.~Yu, E.~Christakopoulou, and A.~Gupta.
\newblock Identifying decision makers from professional social networks.
\newblock KDD '16.

\bibitem{geminix_kdd}
Y.~Zhou, S.~Mishra, J.~Gligorijevic, T.~Bhatia, and N.~Bhamidipati.
\newblock Understanding consumer journey using attention based recurrent neural
  networks.
\newblock {\em KDD}, 2019.

\end{thebibliography}
\end{document}